\documentclass[%
 aip,
 amsmath,amssymb,
 preprint,%
]{revtex4-1}

\usepackage{graphicx}

\DeclareMathOperator\erfc{erfc}

\usepackage{amssymb}
\usepackage{amsthm}
\usepackage{mathtools}

\begin{document}

\preprint{AIP/123-QED}

\title[]{Super-resolution energy spectra from neutron \mbox{direct-geometry} spectrometers}

\author{Fahima Islam}
\thanks{Both authors contributed equally to this manuscript}
\author{Jiao Y. Y. Lin}%
 \email{linjiao@ornl.gov}
\thanks{Both authors contributed equally to this manuscript}
\affiliation{ 
Neutron Scattering Division, Oak Ridge National Laboratory, Oak Ridge, Tennessee 37831, USA
}%
\author{Richard Archibald}
\affiliation{ 
Computer Science and Mathematics Division, 
Oak Ridge National Laboratory, Oak Ridge, Tennessee 37831, USA
}
\author{Douglas L. Abernathy}
\affiliation{ 
Neutron Scattering Division, Oak Ridge National Laboratory, Oak Ridge, Tennessee 37831, USA
}
\author{Iyad Al-Qasir}
\affiliation{ 
Department of Mechancal and Nuclear Engineering,
University of Sharjah,
Sharjah 27272
United Arab Emirates}
\author{Anne A. Campbell}
\affiliation{
Materials Science \& Technology Division,
Oak Ridge National Laboratory, Oak Ridge, Tennessee 37831, USA
}
\author{Matthew B. Stone}
\author{Garrett E. Granroth}
 \email{granrothge@ornl.gov}
\affiliation{ 
Neutron Scattering Division, Oak Ridge National Laboratory, Oak Ridge, Tennessee 37831, USA
}

\date{\today}

\begin{abstract}
Neutron direct-geometry time-of-flight chopper spectroscopy is instrumental in
studying fundamental excitations of vibrational and/or magnetic origin.
We report here that techniques in super-resolution optical imagery (which is in \textit{real}-space)
can be adapted to enhance resolution and reduce noise for a neutron spectroscopy (an instrument for mapping excitations in \textit{reciprocal} space).
The procedure to reconstruct super-resolution energy spectra of phonon density of states relies on
a realization of multi-frame registration,
accurate determination of the energy-dependent point spread function, 
asymmetric nature of instrument resolution broadening, and iterative reconstructions.
Applying these methods to phonon density of states data for a graphite sample demonstrates 
contrast enhancement, noise reduction, and $\sim$5-fold improvement over nominal energy resolution.
The data were collected at three different incident energies measured at the Wide Angular-Range Chopper Spectrometer at the Spallation Neutron Source.
\end{abstract}

\maketitle

\begin{quotation}
\end{quotation}

\section{Introduction}

Inelastic neutron spectroscopy (INS) is a powerful probe of fundamental excitations in solids, including those of vibrational (phonon) or magnetic origins. Phonon measurements play a key role in understanding the physical properties of a solid, such as heat transportation and electrical conductivity  \cite{budai2014, lichen2015, bansal2018momentum, manley2018supersonic}, superconductivity \cite{fong1995phonon, Osborn2001, Rosenkranz2012, RanKeJing}, as well as thermodynamic quantities such as entropy, which influences the phase stability of materials \cite{FULTZ2010, smith2017separating, Kim2015, swan2006vibrational, bogdanoff2002phonon}.
Measurement of changes in phonon frequencies following changes in thermodynamic parameters such as temperature \cite{KreschNickel2007, Delaire2008, Kim2015, Kim2018}, pressure \cite{Klotz2000, Jeong2004}, and chemical composition \cite{RanKeJing} provides important data for understanding the origin of anomalies in phonon behavior, such as anharmonicity.
Measuring the phonon density of states (DOS), a reduced representation of vibrational property of condensed matter, is usually the first step in determining the phonon properties of a material experimentally (See, e.g., \cite{Kim2015, Kim2018}). 

Neutron direct geometry spectrometers (DGS) are important and convenient tools for measuring phonon DOS of a powder sample.
However, resolution functions in DGS measurements of phonon DOS are well known to be cumbersome to model \cite{loong1987resolution}.
The resolution function of the ARCS instrument \cite{abernathy2012design}, a typical DGS instrument, is not a simple Gaussian function,
as usually assumed (justifiably) in most optical imaging techniques. 
For it and all single chopper DGS instruments at a spallation source, the resolution function is asymmetric.  
Such asymmetry arises from the moderation process peculiar to neutron production in spallation neutron sources \cite{ikeda1985}.
This complexity is neglected by most studies so far.

\begin{figure}
\centering
    \includegraphics[width=\linewidth]{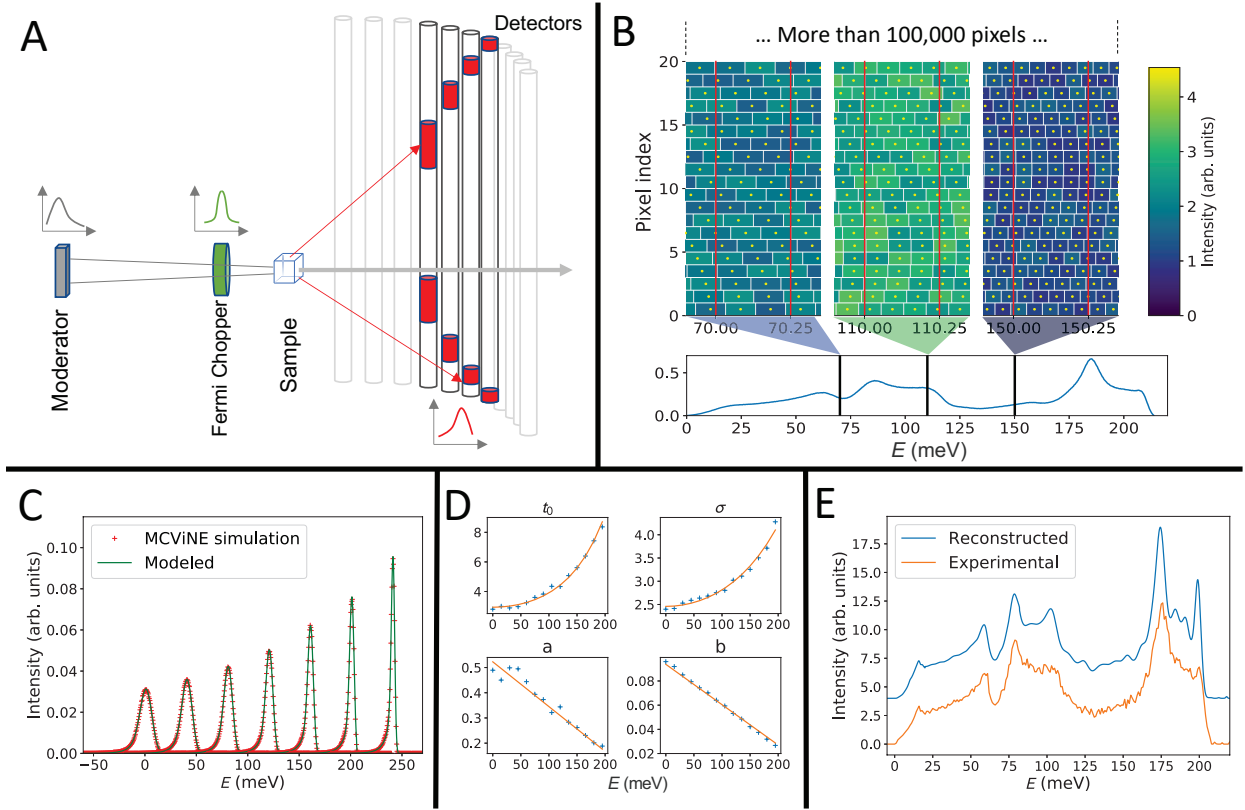}
\caption{\linespread{1.3}\selectfont{}
\footnotesize Principles of super resolution for phonon density of states (SRDOS) measured at neutron direct-geometry spectrometers.
(A) Schematic of a DGS instrument. 
At each pixel a time-of-flight spectrum is measured, which can be converted to an energy spectrum. 
In the experiment described here, each pixel independently measures the phonon DOS. 
The cylindrical layout of detector tubes means the pixels at an equivalent scattering angle are at varying distances from the sample.
The scattering at a specific scattering angle (indicated in red) may include several pixels as it is an intersection of the cone of scattering with the detector.
(B) Illustration of sub-$E$-bin shifts among pixels in an ARCS energy spectrum at $E_i=300$meV.
A phonon DOS energy spectrum is shown at the bottom.
Each value in the energy spectrum is combined from data in many detector pixels.
Zoom-in views of three small energy transfer ranges at 70, 110, and 150 meV are shown on the top.
In the zoom-in views, the vertical axis is pixel index. Only 20 out of $>100,000$ pixels are presented for illustration purposes.
The horizontal axis is energy transfer $E$.
The regular $E$ bin size is assumed to be 0.25meV, and the red vertical lines across all rows show the regular $E$ bin boundaries.
At each row for a particular pixel, colored blocks represent energy bins corresponding to 0.1$\mu$s TOF bins, denoted as $E_{TOF}$ bins.
Sub-$E$-bin shifts exist due to variation of sample-pixel distance, and are obvious from one row to another.
The centers of $E_{TOF}$ bins are marked yellow dots.
 (C) Energy-transfer-dependent resolution at $E_i=300$\,meV for the ARCS instrument at SNS. 
The crosses are MCViNE simulation results, and the solid lines are from model fits. 
(D) Fitted Ikeda-Carpenter-based model parameters for the resolution functions in (C)
as functions of energy transfer. 
Here $t_0$ and $\sigma$ are in microseconds, $a$ and $b$ are unitless.
(E) Reconstruction example. The orange curve is experimental data. The blue curve is offset vertically for presentation and is a reconstruction displaying sharper, less noisy DOS curve.
}
\label{fig:workflow}
\end{figure}

Another complexity associated with the resolution function of DGS instruments is it varies as a function of neutron energy transfer ($E$). 
As $E$ increases the energy resolution broadening decreases (Figure \ref{fig:workflow}C).
When the phonon energy range is not large, resolution variation is small enough so that studies can be done by comparing and/or fitting models 
(for example, Born–von K\'{a}rm\'{a}n models) and DFT calculations to data, involving a convolution with simplified resolution functions \cite{loong1998phonon, Osborn2001, KreschNickel2007, Kim2015}, 
ignoring finer features in data blurred by resolution broadening and statistical noise.
However, for a material with a wide range of phonon energies, the resolution of a DGS instrument at lower energy transfers is unsatisfactory for higher incident energy (Figure \ref{fig:workflow}C). One way to alleviate this problem is to measure the phonon spectrum using multiple incident energies. 
The highest $E_i$ dataset has the largest dynamic range, $r\cdot E_i < E < E_i$ ($r$ is a number that is approximately the relative resolution of the instrument at the elastic line, where there is no energy transfer. For example, for ARCS at SNS \cite{mason2006spallation}, $r\approx 3\%-5\%$), and covers almost the full phonon DOS spectrum, but the resolution of the lower energy transfers are not optimal. By using  a lower $E_i$,  the phonon DOS is measured over a smaller dynamic range, but with finer energy resolution. All DOSes are stitched together to obtain one DOS curve. This is done by replacing the low $E$ portion of the DOS from the higher $E_i$ measurement with the partial DOS obtained from the lower $E_i$ measurement. 
The resolution of the final stitched phonon DOS, however,
is still uneven across the energy range.

Deconvolution techniques have been demonstrated for a few cases in neutron spectra \cite{pynn1990optimization, weese1996deconvolution}, but have not been widely applied. 
The usual objection to deconvolution is that it "generates information from void". 
As we will show below, however, the information content of the DGS data is actually not fully utilized. 
This is particularly true for the modern DGS instruments on spallation sources with finely pixelated cylindrical detector banks such as Merlin, ARCS, SEQUOIA, CNCS, LET, 4-SEASONS, and HRC \cite{BEWLEY20061029,abernathy2012design,Granroth2010,EhlersCNCS,BEWLEY2011128,4seasons,ITOH201190}.
Furthermore, advances in modeling the instrumental resolution function and in reconstruction have only become available recently. This contribution will show how these advances can use the full information content.

Specifically, we demonstrate a super-resolution technique for spectroscopy that builds upon
techniques used in super-resolution imagery.
These techniques work when 1) the sampling frequency in the measurement is higher than that of the nominal resolution; 2) sub-bin shifts (similar to sub-pixel displacements in multi-frame imaging super-resolution) exist for multiple measurements  (for a DGS instrument, one measurement per detector pixel) of the same data; 3) the point spread function has a sharp feature (contains high frequency components).
When these conditions are met, the spectra from multiple measurements can be fused together, and an iterative reconstruction can be used to obtain data in finer resolution.

\section{Methods}

\subsection{Fusion} 

Super-resolution (SR) techniques and statistical methods have seen wide application in optical imaging techniques such as surveillance, forensic science, remote-sensing, and microscopy \cite{zalevsky2004optical, capel2004, cristani2004distilling, li2008universal, maintz1998survey, narayanan2007computationally, park2003super}. 
A typical multi-frame imaging SR procedure \cite{park2003super}
consists of 
an image registration step \cite{irani1991mfsr} (or "fusion"),
followed by a deconvolution of the point spread function.
The fusion step essentially improves the sampling frequency of the "fused" image by combining multiple images of lower resolution
shifted by subpixel displacements, as often seen in, for example, video footage of moving objects.
This important step was not examined in the previous deconvolution work \cite{pynn1990optimization, weese1996deconvolution}.
In direct-geometry spectrometers (see a schematic at Figure \ref{fig:workflow}A),
the spectrum is measured in terms of energy transfer
$E$, obtained by conversion from time-of-flight data: $E=E_i-E_f=E_i-\frac{1}{2}m(\frac{L_{sp}}{t-t_i})^2$, where $E_i$, $E_f$ are initial and final energies of a neutron,
$L_{sp}$ is sample to detector pixel distance, $t$ is the total time of flight (TOF) measured,
$t_i$ is the time-of-flight from neutron moderator to sample.
For direct geometry spectrometers, $E_i$ and $t_i$ are fixed for an experiment.
The time-of-flight at all detector pixels is clocked synchronously to a $0.1 \mu s$ precision.
Therefore as events are accumulated into TOF bins during reduction, the minimum bin size is $0.1 \mu s$ at SNS.
 As detector pixels are at varying distances from the sample (Figure \ref{fig:workflow}A), the constant time binning corresponds to an energy binning that varies across  detector pixels
(Figure \ref{fig:workflow}B).
The typical "event-mode" DGS reduction procedure accumulates events in one array of energy bins.
Therefore the data are fused together similar to multi-frame SR imagery (Figure \ref{fig:workflow}B). 
One obvious difference from multi-frame SR imagery is that for the current work, the data is 1D (energy axis only). More non-trivial differences exist.
First, the conversion from TOF to energy transfer $E$ is not a linear transformation (e.g. affine transformation) as often used in optical SR image registration. 
Further, in contrast to SR imagery, the initial bins (TOF) are usually very fine.
When they are transformed to energy bins, they are usually much smaller than nominal energy resolution.
The conventional wisdom in the DGS data reduction procedure is to use an $E$ bin size that is a fraction of the nominal resolution of the instrument. For example, for ARCS, a $E$ bin size 
($\Delta E$) at 1\% of the incident energy is often used while the nominal resolution is 3-5\%. 
For $E_i=300$meV, $\Delta E=3$meV. Therefore, an $E$ bin would typically encompass many TOF bins. At $E=150$meV, $\Delta E=3$meV corresponds to $\sim$57 TOF bins. However, if we ever want to improve resolution, a higher  sampling rate for $E$ is needed. $\Delta E=0.25$meV would corresponds to $\sim$4 TOF bins. As the transformation is nonlinear, one energy bin almost always corresponds to a non-integer number of TOF bins. 
For example, as shown in Figure \ref{fig:workflow}B, for $E$ near 110meV, one 0.25meV energy bin corresponds to 3 or 4 TOF bins, depending on pixel; near 150meV, one energy bin corresponds to 4 or 5 TOF bins; near 70meV, 2 or 3 TOF bins.
The current event-mode reduction assumes the events happen at exact time-of-flights. As events can happen near the boundaries of energy bins, there is always an error accumulating counts into energy bins for a single detector pixel, even if we perform a measurement for infinite time. 
However, because the variations in sample-pixel distances result in variations of energy bins that have "sub-$E$-bin" components, this kind of error is minimized as data from all pixels are combined, 
and we can safely use an $E$ bin size similar to those corresponding to the 0.1 microsecond TOF bin. This benefit is enhanced by the cylindrical configuration of the detector array. 
With this realization, 
our first modification of the data reduction was to increase the sampling frequency
in the energy axis to at least 3 times the usual binning frequency, or $\sim$15 times finer than the nominal resolution.
The high sampling rate in $E$ provides a solid foundation for the next processing steps in achieving super-resolution.

\begin{figure}
\centering
 \begin{minipage}[c]{0.55\textwidth}
    \includegraphics[width=\textwidth]{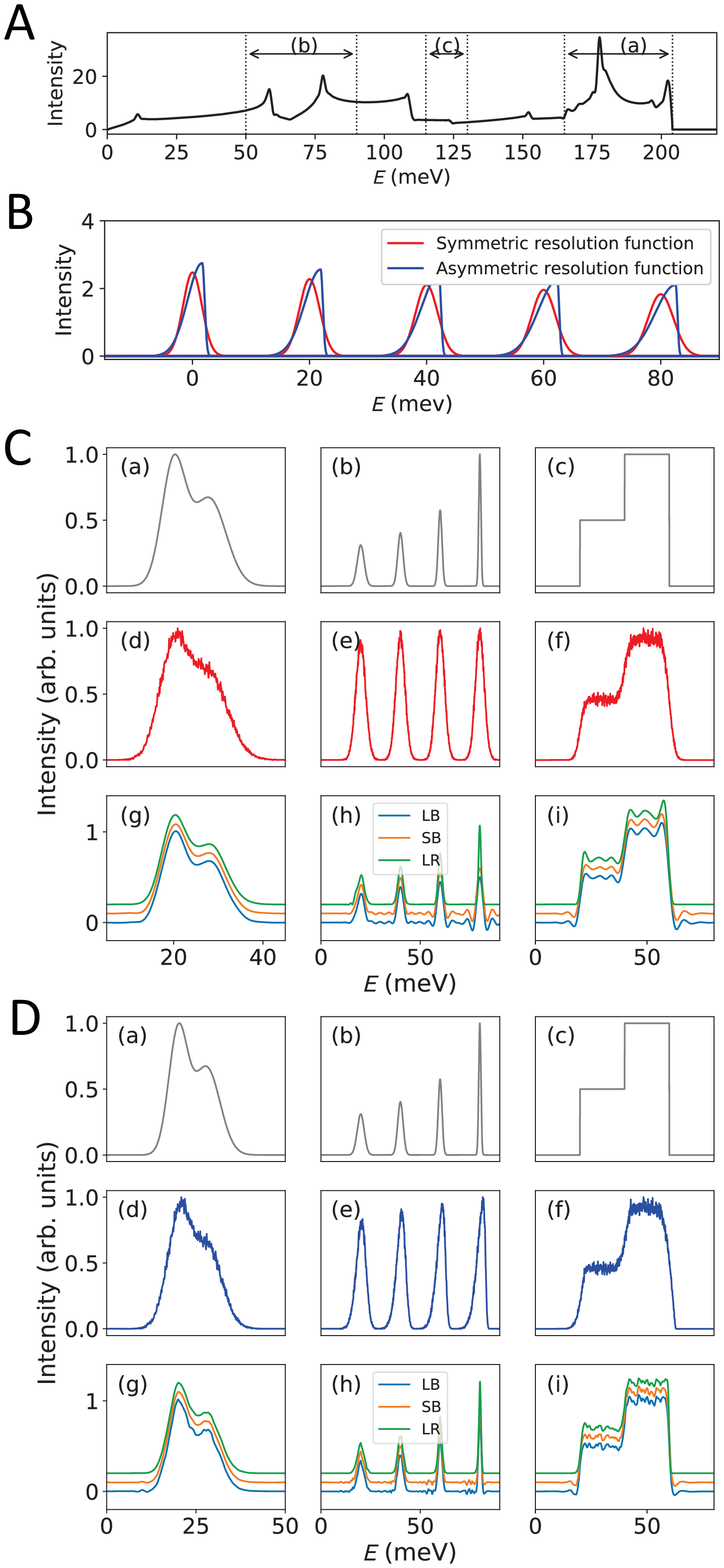}
  \end{minipage}
  \begin{minipage}[c]{0.42\textwidth}
    \caption{\label{fig:synthetic}
\linespread{1.1}\selectfont{} \small Reconstruction test of synthetic data with representative features.
    (A) Phonon DOS of graphite calculated from a DFT simulation.
    This DOS curve shows some common features of a phonon DOS.
    (a) overlapping peaks; 
    (b) multiple separated peaks; (c) sharp, step-like changes in intensity.
    (B) \textbf{Gaussian} and \textbf{asymmetric} resolution functions used in (C) and (D). At each energy both functions have the same FWHM. The asymmetric resolution functions have \textbf{sharp edges}.
    The peak of an asymmetric resolution function looks shifted to the side compared to a Gaussian one while their centers of mass overlap.
    (C) Reconstruction of synthetic DOS curves with \textbf{Gaussian}-smearing and Poisson noise. 
    The original signals in the first row represent typical DOS curve features 
    (a) overlapping peaks (b) separated peaks with varying widths (c) step functions.
    They are convolved with an energy dependent resolution function and a Poisson noise is superimposed 
    resulting in the test signals in the second row. 
    Reconstruction with the candidate algorithms results in the reconstructed signals in the third row, 
    where signals from different reconstruction algorithms have been shifted along the 
    intensity axis for better visualization. 
    (D) Reconstruction of synthetic DOS curves with \textbf{asymmetric}-resolution-smearing and Poisson noise. 
    The plots here are similarly structured as in (C).
    The difference starts from the second row: the synthetic test dataset is created
    by convolving the signals in the first row with an energy-dependent asymmetic
    resolution function and adding Poisson noise.
    The third row shows the reconstructed curves from data in the second row.
    }
  \end{minipage}
\end{figure}

\subsection{Point spread function}

Our next step is to find an accurate model of the "point spread function" (PSF)
for a DGS energy spectrum.
Previously C$_4$H$_2$I$_2$S samples were used to measure 
the inelastic resolution functions experimentally at several energy transfers \cite{abernathy2012design}.
However as such calibrants rely on the molecular structure, they are not tunable and thus modeling must be used to obtain the energy-dependent resolution function at
an arbitrary energy transfer.
Figure \ref{fig:workflow}A show a simple schematic of a DGS instrument. 
One major cause of resolution  broadening in energy is the line shape of neutron spectra emitted from the moderator, which is asymmetric.
As the line shape is a function of time and energy (velocity), filtering by the Fermi chopper selects a narrow band of time and produces a symmetric distribution.  As the neutrons of different speed disperse while traveling to the detector,  the shape becomes asymmetric again but with the tail now at the short times. So the chopper acts like a pin-hole in time of flight \cite{LOONG1987381}.

More factors influence the instrument broadening,
including the divergence of the neutron beam formed through neutron guides, the sample shape and size, and the detector geometry.
Monte Carlo neutron ray tracing simulations with the MCViNE package \cite{jlin2016} can capture these details, and are used to calculate the PSF functions by taking relevant factors into account: modeling the neutron beam, the sample with a $\delta$-function scattering kernel with a geometric shape and scattering cross section same as the real sample, and the detector system in high accuracy \cite{jlin2016, abernathy2012design, diallo2016ARCSres, lin2019energy}.
In Figure \ref{fig:workflow}C  examples of MCViNE-simulated,
energy-dependent energy resolution functions are presented.

In order to obtain resolution functions for energy transfers
across the dynamical range of the measurement, 
it is possible to use MCViNE to simulate at each energy bin a point spread 
function, but such computations are demanding in computing resources.
We chose to first compute the resolution function using MCViNE at a few energy transfers and then fit each of these resolution functions to a revised \cite{loong1993} Ikeda-Carpenter function \cite{ikeda1985} with parameters $a$, $b$, $\sigma$, and $t_0$, depending on energy transfer
(Figure \ref{fig:workflow}D).
These parameters were found to vary slowly across the energy range
of interest, 
and can be interpolated to obtain a PSF at any point along the energy axis.
One thing to remember is, since the phonon DOS may be obtained from
experiments measured using multiple incident energies,
this procedure of simulation, modeling, and interpolation
is required for all incident energies.

An example of the model fitting can be seen in 
Figure \ref{fig:workflow}C.
There, the crosses are MCViNE simulation results of PSF functions,
and the solid lines are from fitted models, which agree very
well with the simulation data points.
An example of the fitted parameters as functions of energy transfer is presented in Figure \ref{fig:workflow}D.
Parameters $a$ and $b$ follow nearly linear relation w.r.t. $E$,
while $t_0$ and $\sigma$ increases quickly when $E$ gets closer
to the incident energy, as expected. \cite{ikeda1985}

\subsection{Reconstruction}

In the following we discuss the methods used 
to reconstruct a one-dimensional (1D) "super-resolution" 
signal from a DGS measurement with instrument broadening and noise.
There are a few reports of the application of deconvolution methods on neutron spectra.  
Sivia, Silver, and Pynn showed that Maximum-entropy method was applicable to deconvolve the resolution function from neutron spectra 
 \cite{pynn1990optimization}, and the asymmetric resolution
function in pulsed source TOF spectrometers is beneficial in the
deconvolution process, due to higher frequency components of the sharp edge
than  symmetric, gaussian-like resolution functions of same full-width-at-half-maximum (FWHM).
Weese et al. later reported that it was possible to deconvolve quasi-elastic neutron spectra using a Tikhonov regularization method \cite{weese1996deconvolution}.
In both efforts, the resolution function was assumed to be independent of energy transfer.

A spectrum measured at a DGS instrument can be in general written in terms of the resolution function and the true, resolution-free spectrum (we call it "ground-truth") as

\begin{equation} \label{eq:f=int_Ru}
    f(E) = \int R(E, E') u(E') dE' + n(E)
\end{equation}
where $f(E)$ is the measured spectrum, 
$R(E, E')$ is the energy-dependent resolution function at energy transfer $E$,
$u(E)$ is the ground-truth spectrum,
and $n(E)$ is the noise term.
A measurement is always discretized. Therefore, 
a discretized version of Eq. \ref{eq:f=int_Ru} is needed:

\begin{equation} \label{eq:f=Ru}
    f = R u + n
\end{equation}
where $f$ is 1D data array for the measurement, $R$ is the 2D resolution matrix, $n$ is the noise array, and $u$ is the ground truth array.
This problem of obtaining $u$ from $f$ becomes ill-posed because of the noise and uncertainty in the PSF.
Regularization techniques are usually employed
to constrain the inverse problems like this to enforce
some a priori knowledge about the ground truth $u$.
The purpose of the reconstruction is:
given the experimental
data $f$ and resolution function $R$,
to find an estimate of the ground truth $u$ that
has improved resolution, by using signal-processing methods.
Many image deblurring methods exist that can be leveraged for this purpose.
We have adapted Lucy-Richardson (LR), one of the earliest deconvolution methods \cite{richardson1972, lucy1974},
and more recent Linearized Bregman (LB) \cite{yin2008LB, cai2009linearized, cai2009convergence}, 
and Split Bregman (SB) \cite{goldsteinSplit, Split} reconstruction methods
to the 1D datasets and taken into account the fact
that the resolution function is energy-dependent.
More mathematical and algorithmic details of the reconstruction techniques based on these three
methods can be found in Appendix \ref{sec:reconmethods}.

\begin{figure}
\centering
    \includegraphics[width=.7\linewidth]{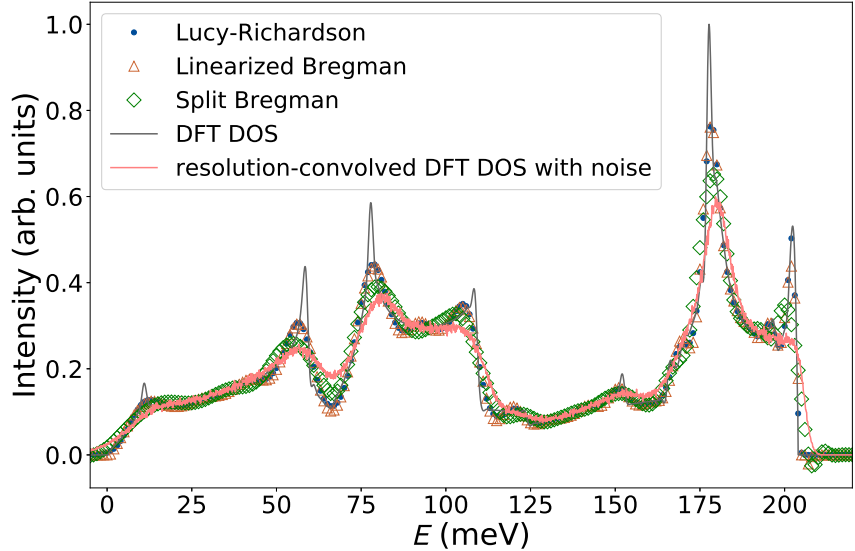}
    \caption{
    Reconstruction of synthetic graphite phonon DOS. 
    The black curve is the graphite phonon DOS calculated from a DFT simulation.
    It is then convolved with ARCS instrument resolution function at $E_i$=300\,meV and then Poisson noise is added to produce the red curve. Other datasets are reconstructed DOS from the red curve using the corresponding algorithm.
}
  \label{fig:DFT-with-noise}
\end{figure}

\section{Results}
\subsection{Synthetic data}

Figure \ref{fig:synthetic}A shows three common features in a phonon DOS curve:
overlapping peaks, multiple separated peaks \footnote{In DGS spectra, single peak widths have important meaning as measure of life-times of excitations},
and step functions. 
The reconstruction methods will help recover these features from noisy and smeared (convolved with resolution ) DOS curves.
The performance of the reconstruction methods are first tested against
a dataset consisting of three synthetic DOS curves with these distinct, representative phonon DOS features,
and the results are presented in Figure \ref{fig:synthetic}C and \ref{fig:synthetic}D.
Shown In Figure \ref{fig:synthetic}C  are test results
on synthetic datasets created by convolving common features with \textbf{symmetric}, Gaussian 
resolution functions, and then adding Poisson noise.
The first column is for the overlapping peaks. Since both peaks are relatively broad, the reconstructed data in panel (g) resemble  well the original data in panel (a), regardless of the reconstruction algorithm.
The second column is for multiple peaks with different widths.
All algorithms perform well when the peak width is large.
For sharper peaks, both the Linearized Bregman algorithm and the Split Bregman algorithm
show oscillations at low intensity, while the Lucy-Richardson method performs well
throughout the range.
The third column is for step functions.
Reconstruction results from all algorithms show some oscillations on top of the plateaus.

In direct geometry spectrometers at spallation sources, 
an important property of the instrument is that usually its energy resolution is \textbf{asymmetric}.
The following test shows how this property can be taken advantage of.
Shown in Figure \ref{fig:synthetic}D are test results
on synthetic datasets created by convolving common features with \textbf{asymmetric} 
resolution functions, and then adding Poisson noise.
The results are similar for broader peaks in column one and two.
Reconstructions in column two and three for all algorithms show less oscillation
than in Figure \ref{fig:synthetic}C.
This can be attributed to the \textbf{sharper edge} of the asymmetric resolution function
used here. 
The effects of the sharp edge of the resolution function in spallation neutron source data was originally discussed in \cite{pynn1990optimization}.

In summary,  the reconstruction algorithms work better for the synthetic 
overlapping-peaks dataset in the first column, than the synthetic 
multiple-peaks and step-function datasets in the second and third columns.
The underlying reason, however, is that the multiple-peaks dataset and the step-function datasets consist of higher-frequency components such as sharper peaks and edges, 
which are harder to resolve.

We then test the reconstruction algorithms with a synthetic DOS curve for graphite.
This dataset is created by convolving the original DOS calculated from a DFT simulation
with the ARCS instrument resolution function for $E_i=300$\,meV, and adding Poisson noise.
The results are shown in Figure \ref{fig:DFT-with-noise}.
The reconstructed DOS curves show clear improvements in recovering
peaks at 200, 180, and 80\,meV,  and cliffs (step-functions) at 205 and 165\,meV,
compared to the resolution-smeared DOS curve.
The Split Bregman method does not work as well for peaks at 200 and 180 meV, and is not the recommended method.
More quantitative analysis of the reconstruction efficacy of the algorithms can be found in Appendix \ref{sec:reconefficiency}.

\begin{figure}
\centering
    \includegraphics[width=.8\linewidth]{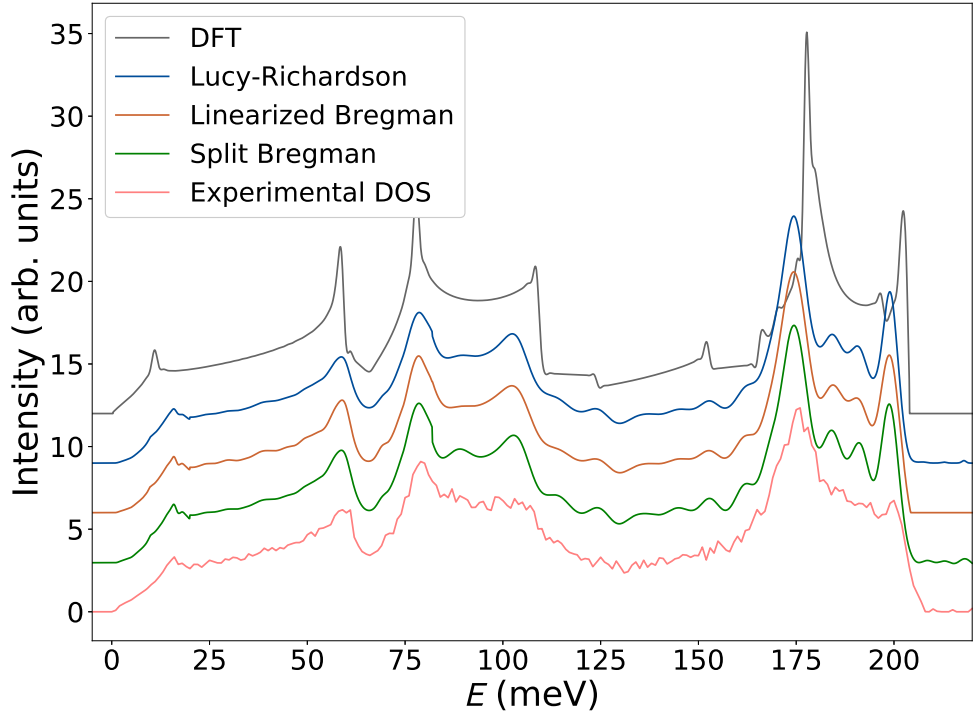}
    \caption{
    Comparison between reconstruction results using different algorithms (three curves in the middle) from the experimental DOS (red curve at the bottom). DOS calculated from the DFT simulation is also presented (black curve).
    }
  \label{fig:RealGraphiteDOS}
\end{figure}

\subsection{Experimental phonon DOS}
We apply the SRDOS workflow to a phonon density 
of states dataset for a graphite sample measured at the ARCS instrument at SNS.
Three different $E_i$s: 30\,meV, 130\,meV and 300\,meV were used for the measurement and the resultant data were reduced to the Phonon DOS using the standard procedure. 
We then simulate and interpolate the resolution functions as explained earlier (Figure \ref{fig:workflow}).
A reconstruction algorithm is then applied to the three phonon DOS spectra
measured using different $E_i$s, taking into account
the energy-dependent, asymmetric resolution functions 
to obtain three super-resolution DOS curves,
which are then stitched together to obtain a full DOS curve.
The results from three reconstruction algorithms are plotted in Figure \ref{fig:RealGraphiteDOS},
along with the theoretical phonon DOS obtained from 
DFT simulation and the experimental phonon DOS.
Compared to the experimental DOS at the bottom, the reconstructed DOS curves enhance peaks near 200, 155, and 105\,meV;
sharpen peaks near 180 and 80\,meV, and recover more details in the energy range 120--160\,meV.
The DFT and measurement based phonon DOS curves agree to the level expected by the calculation technique \cite{togo2015first} \footnote{Sample defects may also cause differences.},
demonstrating the power of the SR reconstruction.
Among the three reconstruction algorithms, the Lucy-Richardson method and the Linearized Bregman method
show very similar results, while the Split Bregman reconstruction show an extra bump near 85\,meV, 
which does not exist in the DFT DOS.
The Split Bregman reconstruction also show slightly stronger fluctuation beyond 210\,meV, 
the high energy cutoff for graphite phonons, than the other two reconstruction results.

\section{Conclusion}

Super-resolution phonon DOS spectra were obtained from $S(Q,E)$ measured at ARCS using different incident energies. 
This is done by binning neutron event data in a sampling rate much higher than nominal instrument resolution, taking advantage of "sub-bin" shifts among detector pixels; modeling the resolution function with high accuracy;
taking advantage of high frequency components in sharp edge of the asymmetric DGS resolution function;
and adapting and using reconstruction algorithms from SR imagery.
For example, 
as long as enough counts are collected, the sampling rate for energy transfer can be safely increased to one bin per 0.1meV, 1/100 of nominal energy resolution (10meV) for an $E_i=300$meV measurement at the ARCS instrument, corresponding to $\sim$2 TOF bins of $0.1\mu s$ at $E=150$meV. 
The sampling rate in the $E$ axis can be further improved slightly when TOF data are converted to energy data by splitting events to neighboring bins or by interpolation.
Another important factor is determined to be the sharper edge of the resolution function. For example, the sharper edge at $E=200$meV has a half width of $\sim$2meV, and the effective resolution is smaller and estimated to be $\sim$1.1meV.
It means that the resolution of reconstructed data at $E=200$meV is $\sim$14 times better than the elastic resolution ($\sim$15meV) of the measured $E_i=300$ dataset. With measurements using multiple incident energies, $\sim$5$\times$ resolution improvement across the energy range is readily accessible. 
This technique is limited by the signal-to-noise ratio as other reconstruction techniques, although for our synthetic datasets, noise level less than $\sim$3\% of maximum intensity is found to have little effect in reconstruction results.
The difference of the resolution functions in datasets from different incident energies could be used to further refine the reconstruction results that are measured at multiple incident energies.
This technique reduces the influence of instrument-specific resolution in the measured spectra
and allows researchers to examine data closer to the “ground truth”, 
taking advantage of existing research on image deconvolution and denoising. 
It makes use of information already in the current data, 
without going beyond the Nyquist frequency of the measurement.
Similar techniques can be extended to higher dimensional DGS data, such as 2D $S(Q,E)$ in powder measurements,
and 2D slice and 3D/4D volume data in single crystal $S(\mathbf{Q}, E)$ measurements, and possibly to other neutron scattering techniques, provided the three conditions are met: 1) a measurement sampling frequency higher than that of the nominal resolution; 2) sub-bin shifts for multiple measurements of the same data; 3) the point spread function has a sharp feature.

\begin{acknowledgments}
The authors thank J. M. Borreguero Calvo, A. T. Savici, T. E. Proffen, M. K. Stoyanov, and T. R. Charlton for fruitful discussions. The graphite sample of grade G347A was funded by Strategic Planning Partnership between ORNL and Tokai Carbon Co., Ltd. (NFE-09-377 02345) with the U.S. Department of Energy. 
The DFT simulations were performed at the High Performance Facility in University of Sharjah.
This work is supported by the Department of Energy, Laboratory Directed Research and Development SEED funding, under contract  DE-AC05-00OR22725.
This research used resources at the Spallation Neutron Source, a DOE Office of Science User Facility operated by the Oak Ridge National Laboratory.
\end{acknowledgments}

\appendix

\section{Neutron scattering experiment and reduction to phonon density of states}\label{exp_red}
The wide Angular-Range Chopper Spectrometer (ARCS) at the Spallation Neutron Source (SNS) at Oak Ridge National Laboratory \cite{abernathy2012design}  was used to measure the phonon density of states of the nuclear graphite grade G347A with density of 1.85 g/cm3 at room temperature. The incident neutron energies $E_i$=30, 130 and 300 meV were used to cover the whole energy transfer range ($\sim$200 meV) in graphite. 
This measurement is a part of an extended study of the phonon DOS
of nuclear graphite samples after irradiation by fast neutrons \cite{campbell2016property}.
The full details of these measurements will be
published elsewhere \cite{IyadGraphite2019}.
The neutorn scattering data at 3 incident energies were collected as events and saved as NeXus \cite{nexus2015} files,
which are reduced by using Mantid \cite{mantid} to $S(Q,E)$ spectra.
At each temperature the multiphonon \cite{lin2018multiphonon} package is used to iteratively
improve a trial phonon DOS spectrum that fits the $S(Q,E)$ spectrum best
in the incoherent scattering approximation, 
taking multiphonon contribution into account.
The three phonon DOS spectra can be "stitched" together by using
the multiphonon \cite{lin2018multiphonon} package to form one DOS spectrum.
The stitching starts with the phonon DOS measured with the largest incident energy, $g_{E_i=300}(E)$
covering the whole DOS spectrum.
Moving on, the DOS curve is updated
by replacing the low $E$ portion of the DOS $g_{E_i=300}(E)$ with the partial DOS
obtained from the $E_i=$130meV measurement, to form $g_{E_i=130,300}(E)$.
Similarly we can perform the update for $E_i=$30meV measurement.

\section{Ab initio calculations}
The vibrational properties of graphite are calculated using first-principles calculations as implemented by the VASP code \cite{kresse1996efficient,kresse1996efficiency}. 
A 6x6x1 supercell with 144 atoms was used to calculate the Hellman-Feynman forces.  
For the exchange-correlation functional, the generalized gradient approximation (GGA) of Perdew, Bruke and Ernzerho  formalism was used \cite{perdew1996generalized}. 
The integration over the Brillouin zone was performed using a 3x3x4 $k$-mesh \cite{monkhorst1976special} and energy cutoff 900 eV. 
The phonopy code \cite{togo2015first} was used to calculate 
the phonon density of states. 

\section{Monte Carlo neutron ray tracing simulations of PSF}
In this work, we used the "dgsres" package \cite{linjyy2017dgsres}
to calculate the resolution function at selected energy transfers
by performing MCViNE simulations \cite{jlin2016}.
One such simulation consists of an incident beam simulation, a simulation of
neutron scattering from a powder-resolution sample 
which scatters
neutrons only to a particular combination of $Q,\;E$,
and a simulation of detection of the scattered neutrons by
the ARCS detector system which generates an event-mode NeXus file.
The simulated data file is then reduced by Mantid \cite{mantid} to obtain
$I(E)$ spectrum, which would be the point-spread-function (PSF) at
the $Q,\;E$ point of choice.
Since the energy resolution is nearly independent of $Q$, we only
calculate the energy-dependent PSF for one $Q$ value.
Due to the shape of the dynamical range measured by a DGS instrument \cite{windsor1981book},
we choose a $Q$ value to allow for calculation of larger $E$ values.

\section{Fitting PSF}

The model \cite{loong1993} we choose to fit the resolution function is based on the Ikeda Carpenter function \cite{ikeda1985}.
In this model, the time distribution of neutron counts at detector is written as 
\begin{equation}
    \begin{split}
    C(t) = \frac {A}{l}\frac{1}{\sqrt{2\pi}\sigma} \bigg\{ &(1-R)a^2 C_2(a,t) \\   
    &+\frac{Ra^2b}{(a-b)^3} \big[2C_0(b,t)-((a-b)^2C_2(a,t) + 2(a-b)C_1(a,t)+2C_0(a,t)) \big] \bigg\}
    \end{split}
\end{equation}

\noindent where
\begin{gather}
    C_0(x,t)=\sqrt{\frac{\pi}{2}}\left(\frac{\sigma l}{l_2+l_3}\right)\exp({v_{min}^2-u_{min}^2}) \, \erfc(v_{min})\\
    \begin{split}
    C_1(x,t)=&\left(\frac{\sigma l}{l_2+l_3}\right)^2\exp({v_{min}^2-u_{min}^2})\times[\exp{(-v_{min}^2)}-\sqrt{\pi}v_{min}\erfc(v_{min})]\\
    \end{split}\\
    \begin{split}
    C_2(x,t)=&\sqrt{2}\left(\frac{\sigma l}{l_2+l_3}\right)^2\exp({v_{min}^2-u_{min}^2})\times[\sqrt{\pi}(\frac{1}{2}+v_{min}^2)\erfc(v_{min}) -v_{min}\exp({-v_{min}^2})]\\
    \end{split}
\end{gather}
and 
\begin{gather}
    v_{min}=u_{min}+\frac{x}{\sqrt{2}}\Big(\frac{\sigma l}{l_2+l_3}\Big)\\
    u_{min}=\frac{1}{\sqrt{2}\sigma}\Big(\frac{l_1}{l}t-t_0\Big)\\
    \erfc(z)=\frac{2}{\sqrt{\pi}}\int_{z}^{\infty} dx \exp{(-x^2)}
\end{gather}
Here $l_1$, $l_2$, $l_3$ are moderator to Fermi-chopper, Fermi-chopper to sample, and sample to detector distances.
The parameters in this model are $a$, $b$, $R$, which are related
to the parameters in the original Ikeda-Carpenter model \cite{ikeda1985} for the moderator,
and $\sigma$ for broadening caused by factors other than the moderator, including sample and detector,
and $t_0$, a  time offset parameter.
It is then straightforward to transform the time distribution to energy
distribution for comparison with MCViNE simulated PSF.
We have chosen to keep the parameter $R$ to be constant at 0.3. 

\section{Reconstruction methods \label{sec:reconmethods}}
\subsection{Lucy-Richardson method}
Lucy-Richardson (LR) method \cite{richardson1972, lucy1974}
is one of the classical methods
for deconvolution.
When it converges, it converges to the
maximum-likelihood (ML) estimate of the solution
of Eq. 2 of the main manuscript. 
It has found wide applications in deblurring images in,
for example, astronomy \cite{white1994image,tsumuraya1994iterative,starck2002deconvolution}.

The adapted LR method uses the following iteration:

\begin{align}
    u^{k+1} &=& R^{*} \frac{f}{R u^k} \; u^k
\end{align}
where $R^{*}$ is the matrix for the flipped resolution functions.
The initial condition for the iteration is

\begin{equation}
    u^0 = f
\end{equation}

\subsection{Linearized Bregman method}
One of the state of the art methods for restoring noisy and blurry images is the Linearized Bregman method \cite{yin2008LB}.
It is an iterative regularization procedure for solving the basis pursuit problem:

\begin{equation}
    min\{ ||u||_1: Ru=f \}
\end{equation}
where $||u||_1$ is the $L_1$ norm of $u$. It was proved \cite{cai2009linearized, cai2009convergence} that the Linearized Bregman iteration converges to the solution of 

\begin{equation}
\label{eq:minimizatio_L1_and_L2}
    min\{ \mu ||u||_1 + \frac{1}{2\delta} ||u||^2 : Ru=f\}
\end{equation}
where $||.||$ is the $L_2$ norm, and 
$\mu$ is the regularizing parameter. The linearized Bregman iteration which we will adapt, analyze, and use here is generated by

\begin{gather}
    v^{k+1}=v^k-R^T(Ru^k-f)\\
    u^{k+1}=\delta \cdot shrink(v^{k+1}, \frac{1}{\mu})
\end{gather}
where $v^k$ is an auxiliary variable and $\delta$ is the step size of the iteration
The $shrink$ function

\begin{equation}
shrink(x,\mu): = \begin{cases}
x-\mu & \text{if $x > \mu$}\\
0 & \text{if $-\mu\leq x \leq\mu$}\\
x+\mu & \text{if $x <-\mu$}
\end{cases}
\end{equation}
is the soft thresholding algorithm.
The initial condition is $u^0=v^0=0$.

It was proved \cite{cai2009linearized} that the Linearized Bregman iteration converges when
$0 < \delta < \delta_{max} = \frac{2}{||A A^T||}$.
In general a larger $\delta$ value converges
faster than a lower $\delta$ value.
Therefore a $\delta$ value close to $\delta_{max}$
is usually used.

To compute the value of the parameter $\mu$, we chose
to balance the L1-norm and L2-norm regularization terms:
\begin{equation}
\label{eq:LB_calc_mu}
    \mu = \frac{||f||^2}{2\delta ||f||_1}
\end{equation}

The stopping criterion for our denoising algorithm is when the residual is smaller than the error bar:
\begin{equation}
\label{eq:convergence_test}
    || R u - f || < ||\sigma||
\end{equation}
Here $\sigma$ is the errorbars of the measurement $f$.

\subsection{Split Bregman Method}

Split Bregman use the Bregman iteration for minimizing the $L_1$ and $L_2$ terms in \ref{eq:minimizatio_L1_and_L2} separately by decoupling them.
The decoupling is achieved by incorporating an alternative minimization scheme \cite{goldsteinSplit, Split}. 
In this scheme the $L_1$ term in \ref{eq:minimizatio_L1_and_L2} is replaced by a auxiliary variable "$d$" and the equation become:
\begin{equation}
\label{eq:sb_minimization}
    \min\{ ||d||_1 + \mu ||Ru-f||_2^2: d=\phi u\}
\end{equation}
where $\phi$ is the linear operator of $u$. Equation \ref{eq:sb_minimization} is first minimized with respect to $u$ by keeping $d$ fixed and in the next step it is minimized with respect to $d$ while $u$ is kept fixed.
It has been shown that, the alternate minimizing approach reach the convergence with less number of iterations than the conventional approach \cite{SplitConver}.
The split Bregman method to solve the above problem is presented in following:
\begin{equation}
     (u^{k+1}, d^{k+1})=\arg\min_{u,d}\parallel d\parallel_1+\mu\parallel Ru-f\parallel_2^2+\lambda\parallel d-\phi u-b^k\parallel_2^2
\end{equation}
\begin{equation}
\label{eq:SB_b_update}
     b^{k+1}=b^k+\phi u^{k+1}-d^{k+1}
\end{equation}

where $b$ is the bregman vector and updated using \ref{eq:SB_b_update} and $\lambda$ is the regularizing parameter. Here we use gradient descend method and $shrink$ function to solve $u$ and $d$ respectively. Therefore, each iteration of the split Bregman algorithm consists of three steps:
\begin{gather}
    u^{k+1}=u^k-\alpha(\mu R^T(Ru^k-f)+\lambda\nabla^T(\nabla u^k-d^{k}+b^k)) \\
    d^{k+1}=shrink(\nabla u^{k+1}+b^k, \frac{1}{\lambda}) \\
     b^{k+1}=b^k+\nabla u^{k+1}-d^{k+1}
\end{gather}
where $k$ is the iteration index and $\alpha$ is the step size. For the low values of $\alpha$, 
the convergence is slow. 
$\mu$ is caluculated using \ref{eq:LB_calc_mu} and the same principle is used to calculate $\lambda$ from the following equation:
\begin{equation}
    \lambda = \frac{||f||^2}{\mu ||\nabla f||_1}
\end{equation}
The iteration begins with the initial assumption of $u^0=0, d^0=0, b^0=0$ and ends when Equation \ref{eq:convergence_test} is satisfied.

\section{Quantification of reconstruction efficacy \label{sec:reconefficiency}}
\begin{figure}
\centering
    \includegraphics[width=.75\linewidth]{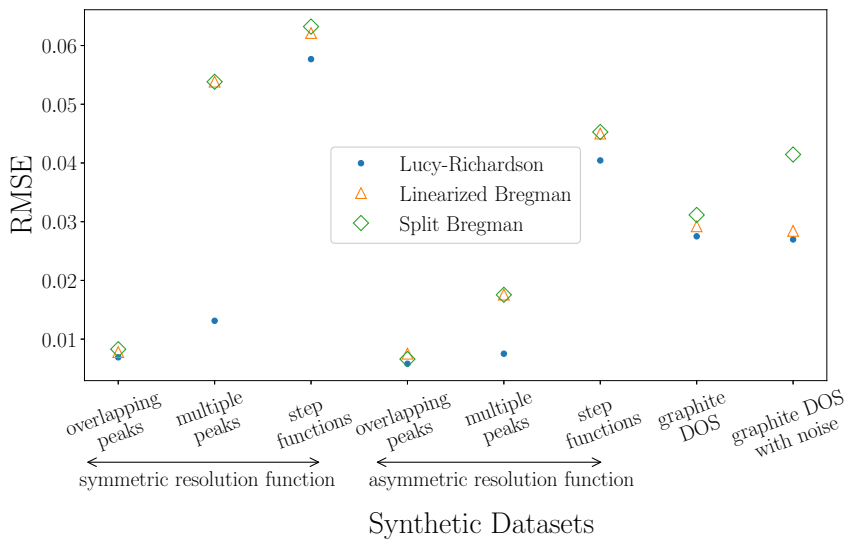}
    \caption{RMSE (see text for definition) for synthetic datasets using three reconstruction algorithms.
    }
  \label{fig:RMSE}
\end{figure}

In order to further quantify the efficacy of the reconstruction methods 
the root mean square error (RMSE) between the reconstructed dataset and the original dataset (ground truth)
is calculated using the following equation:
\begin{equation}
    RMSE = \sqrt{ \frac{\sum(u_r-u_{gt})^2}{L}}
\end{equation}
where, $u_r$, $u_{gt}$ denote the reconstructed data and the ground truth respectively. $L$ is the size of the data.
To make it easier to compare different datasets, each ground truth was scaled so that the maximum intensity is 1.

Figure \ref{fig:RMSE} shows RMSE values for different datasets using different reconstruction algorithms. 
The RMSE values are used to measure how close the reconstructed results resemble the original data, and  
with Figure \ref{fig:RMSE} we can make the same conclusions as what we have observed by eye earlier 
in Figures \ref{fig:synthetic} and \ref{fig:DFT-with-noise}.
It shows that RMSE
is a good measure of reconstruction fidelity.
For example, from Figure \ref{fig:RMSE} it can be seen that 
all three algorithms perform the best for the synthetic overlapping-peaks dataset with small RMSE,
and the disagreement (RMSE) increases for the synthetic multiple-peaks and step-functions datasets,
regardless of the choice of resolution function (symmetric vs asymmetric).
The Lucy-Richardson algorithm apparently performs better than the other two algorithms on the synthetic multiple-peaks dataset.
From Figure \ref{fig:RMSE} we also see that all algorithms perform similarly well for the synthetic graphite DOS without noise,
however the RMSE is in between the RMSE for the synthetic overlapping-peaks dataset and that for the
synthetic step-function dataset, demonstrating a more realistic DOS curve consists of
signals of varying features -- both slowly varying valleys and peaks, as well as sharp peaks and cliffs.
For the synthetic graphite DOS with noise, the split Bregman method yields
a larger RSME than the other two methods.

\begin{figure}
\centering
    \includegraphics[width=.85\linewidth]{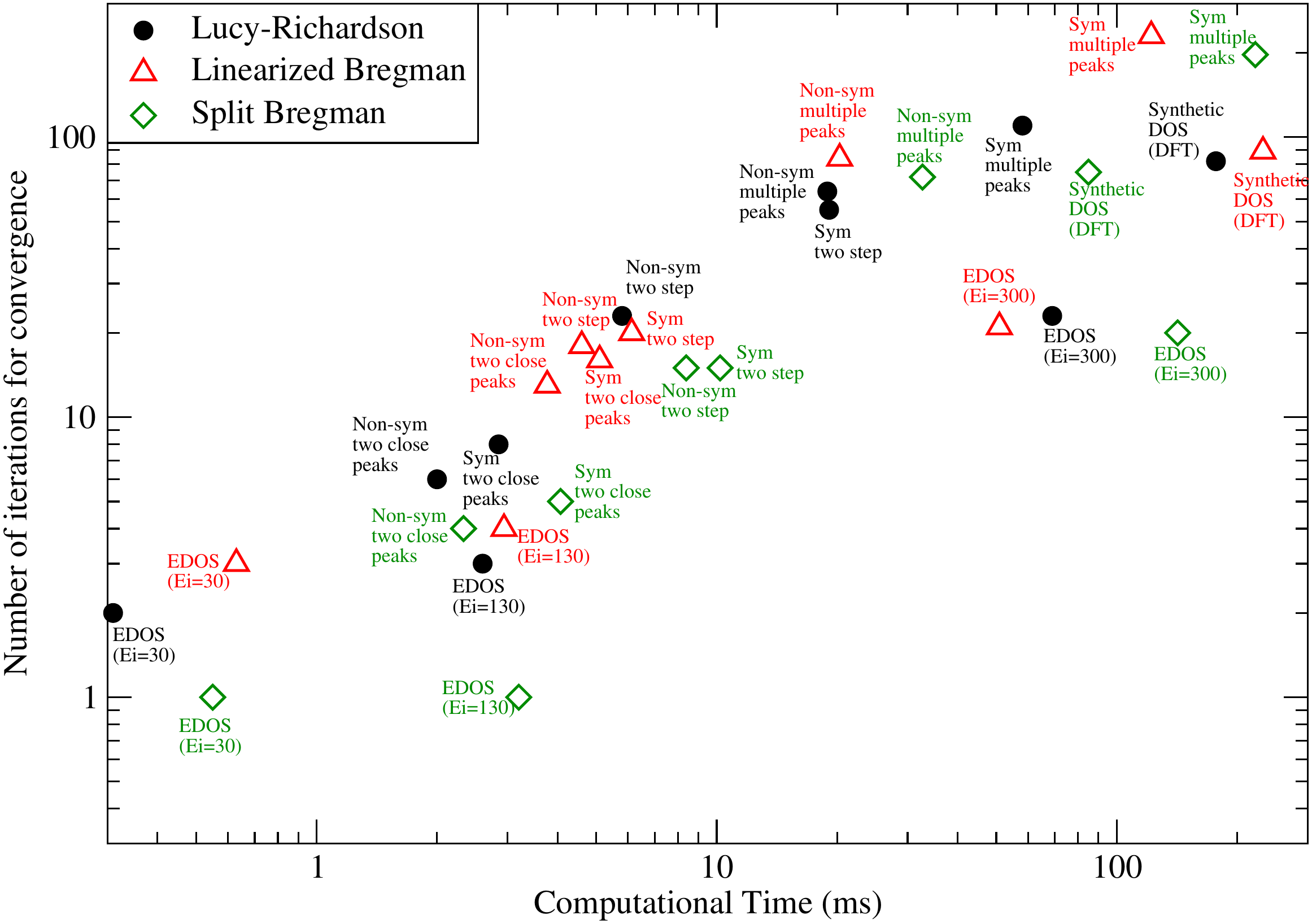}
    \caption{Computational cost for various datasets using three reconstruction algorithms.}
  \label{fig:itervsTime}
\end{figure}

\section{Computation cost}

\begin{table} 
\centering
    \caption{The computational time in ms, the number of iteration required (within parenthesis) 
    and RMSE (bottom) for different 
    synthetic data sets for different reconstruction algorithms. 
    For example, it takes the Lucy-Richardson algorithm 2.85\,ms in 8 iterations to converge, and its RMSE was 0.0069.}
    \label{table:figure_of_merit}
    \includegraphics[width=.85\linewidth]{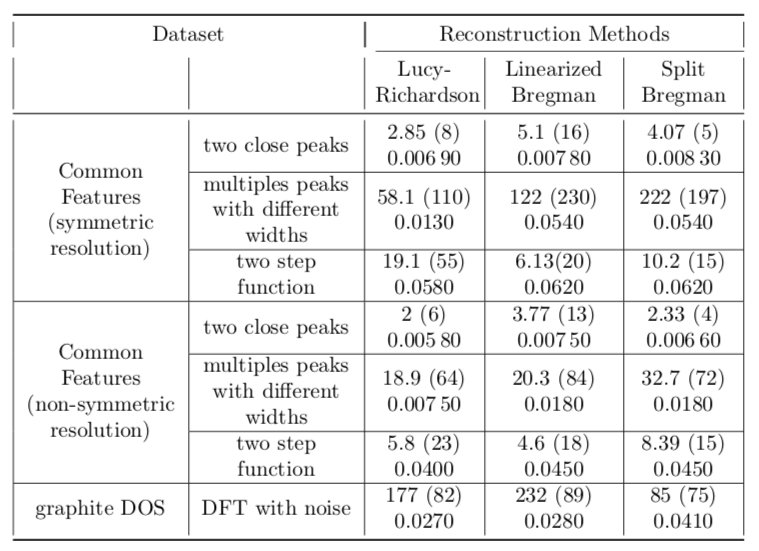}
\end{table}

Figure \ref{fig:itervsTime} and Table \ref{table:figure_of_merit} present the computational costs. 
This figure shows the number of iterations and the computational times 
required to meet the convergence criteria for different datasets. 
The computational time is the "wall time" which can be defined as the actual time for the program to finish from its start time,
and it  is dependent on the machine.
This study was performed on a regular workstation with an Intel Xenon E5-2630 CPU.
Roughly the number of iterations and computation time follow a power law,
and computation time increases with number of iterations.
It is clear from Figure \ref{fig:itervsTime} that in general the Split Bregman method needs
fewer iterations to converge than the other two algorithms,
meaning the Split Bregman method is more efficient in each iteration getting closer to its solution. 
However, the computation time is in general still larger for the Split Bregman method, meaning
the efficiency in each iteration does not convert to overall efficiency.
The Linearized Bregman method and the Lucy-Richardson method shows similar number of iterations and computation time.
All three algorithms are comparatively fast, requiring on the order of milliseconds.

\bibliography{main}

\end{document}